 \documentclass[smallabstract,smallcaptions]{dccpaper}

\usepackage{cite}
\usepackage{graphicx}
\usepackage{url}
\usepackage{amsmath}
\usepackage{amssymb}
\usepackage[caption = false, font=footnotesize]{subfig}
\usepackage{float}
\usepackage{booktabs} 
\usepackage{multirow}
\usepackage[colorlinks=true,linkcolor=blue]{hyperref}%
\usepackage{array}
\usepackage{color}
\usepackage{tabularx}
\usepackage{longtable}
\usepackage{tabu}
\usepackage{pifont}
\usepackage{amssymb}
\usepackage{ragged2e}
\usepackage{tabularx}
\usepackage{longtable}
\usepackage{booktabs}
\usepackage{rotating}
\usepackage[ruled,vlined,linesnumbered]{algorithm2e}

\usepackage{makecell}
\begin{document}

\newcommand\blfootnote[1]{%
  \begingroup
  \renewcommand\thefootnote{}\footnote{#1}%
  \addtocounter{footnote}{-1}%
  \endgroup
}

\title{\textbf{From Scattered Gaussians to Structured Maps: Efficient Gaussian Splatting Coding via Dual-phase Morton Sorting}}

\author{Bolin Chen$^{\natural }$$^{\dag}$$^{\emptyset}$, Shanzhi Yin$^{\ast}$, Ru-Ling Liao$^{\dag}$, Yibo Fan$^{\natural }$  and Yan Ye$^{\dag}$\\[0.5em]
{\small\begin{minipage}{\linewidth}\begin{center}
\begin{tabular}{c}
$^{\natural}$ Fudan University   $^{\dag}$ DAMO Academy, Alibaba Group \\
$^{\emptyset}$ Hupan Lab   $^{\ast}$ City University of Hong Kong
\end{tabular}
\end{center}\end{minipage}}}

\maketitle
\begin{abstract}
3D Gaussian Splatting (3DGS) enables high-fidelity novel view synthesis but suffers from excessive storage and bandwidth requirements due to its unstructured representation. To address this, a projection-based video coding framework has emerged as a leading approach—supported by MPEG’s ongoing standardization—where 3DGS attributes are converted into 2D maps to take advantage of efficient compression using established video codecs such as HEVC and VVC. However, the effectiveness of this approach depends heavily on spatial coherence of the projected video, which current sorting strategies such as PLAS and Morton ordering fail to preserve adequately, either incurring high computational cost or achieving limited correlation retention. To overcome these limitations, we propose a dual-phase Morton spatial sorting algorithm that improves both coding efficiency and processing speed. In the first phase, Morton-based 1D indexing is applied to high-dimensional attributes to enhance spatial locality. The second phase further refines layout continuity through a structured 2D Morton mapping table that enforces spatial adjacency. This hierarchical strategy generates highly regular, block-wise feature maps with strong local correlation, making them well-suited for compression via conventional block-based coding tools. Experimental results show that our method significantly outperforms existing approaches in both compression performance and runtime efficiency, providing a practical and standard-compatible solution for 3DGS data coding.
\blfootnote{This work was supported by the Zhejiang Provincial Postdoctoral Research Project Funding Program (Project No. ZJ2026187).}
\end{abstract}

\vspace{-0.6em}
\section{Introduction}
\vspace{-0.6em}

3D Gaussian Splatting (3DGS)~\cite{kerbl3Dgaussians} has recently emerged as a powerful representation for novel view synthesis, owing to its high rendering quality, fast rasterization, and flexible scene modeling capability. By representing a scene as a collection of anisotropic Gaussian primitives, 3DGS can produce visually convincing rendering results and has shown strong potential in a wide range of applications, including immersive telepresence, augmented reality, virtual reality, and interactive content creation. However, the high fidelity of 3DGS comes at the cost of substantial storage and transmission overhead. A typical 3DGS scene contains a large number of Gaussian primitives, each associated with multiple attributes such as position, scale, rotation, opacity, and spherical harmonic coefficients. Accordingly, the raw representation is highly redundant and costly to transmit, underscoring the need for efficient compression methods, particularly for real-time streaming and resource-constrained mobile applications.

\begin{figure*}[t]
\centering
\vspace{-1cm}
\includegraphics[width=\linewidth]{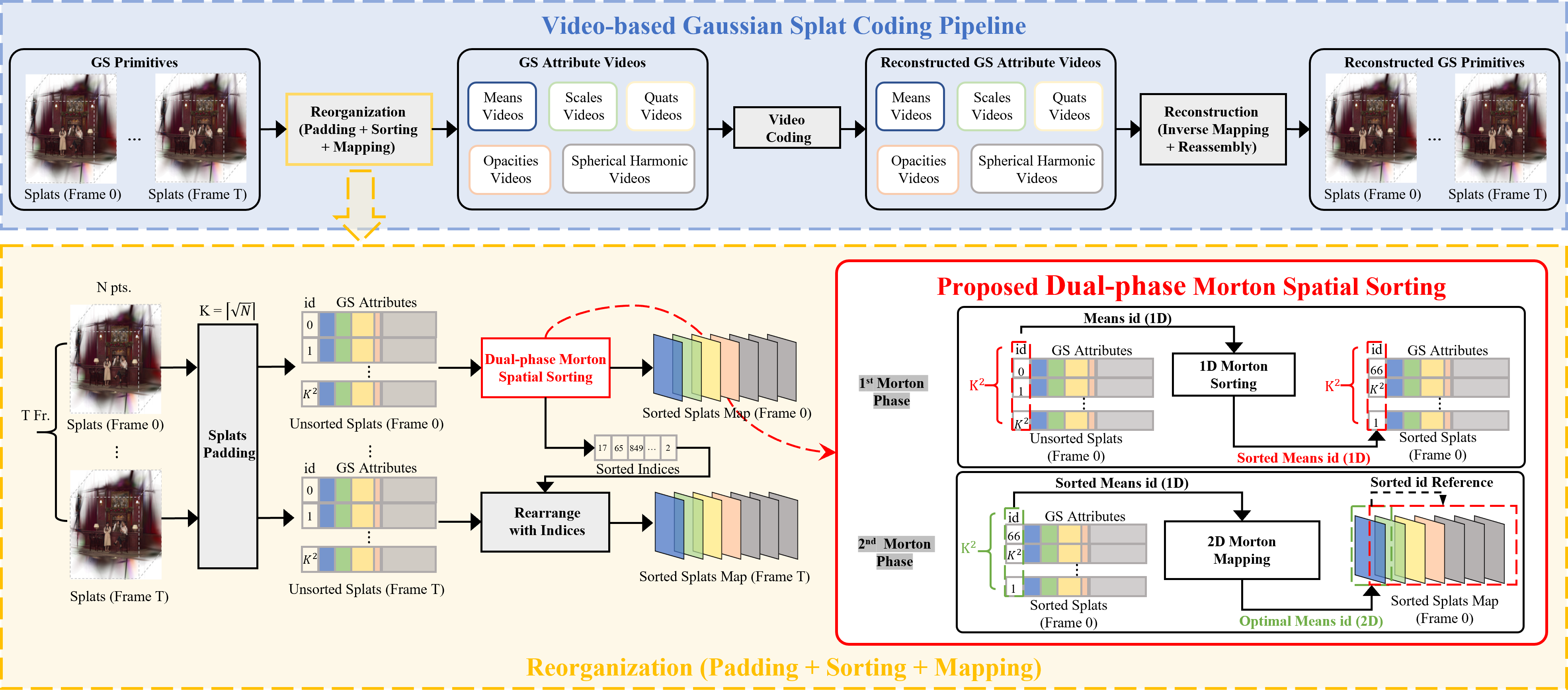}
\caption{Illustration of the video-based Gaussian splats coding pipeline within the proposed dual-phase Morton spatial sorting algorithm.}
\label{framework_detail}
\vspace{-0.4cm}
\end{figure*}

While algorithmic advances continue to accelerate~\cite{3DGSzip2025,11449264}, standardization efforts are turning toward practical coding frameworks for Gaussian primitives. Among them, MPEG’s ongoing work on Gaussian splat coding highlights video-based representation as a promising approach for efficient compression. As illustrated in Figure~\ref{framework_detail}, the core principle is to reorganize per-frame Gaussian attributes into 2D attribute maps and then compress them using mature video coding standards such as High Efficiency Video Coding (HEVC)~\cite{sullivan2012overview} and Versatile Video Coding (VVC)~\cite{bross2021overview}. By transforming irregular 3D Gaussian data into structured 2D arrays, this paradigm allows 3D Gaussians to be coded using conventional codecs to better exploit spatial and temporal redundancies and to benefit from the long-standing optimization of existing video coding toolchains. Under this framework, the sorting and mapping strategy used to arrange Gaussian splats in 2D space becomes a critical component, since the spatial coherence and block regularity in the resulting 2D arrays directly influence the overall coding efficiency.

In video-based 3DGS compression, spatial sorting plays a crucial role in determining how Gaussian attributes are laid out in 2D pixel space. Existing methods mainly include PLAS~\cite{morgenstern2024compact} and Morton-based ordering~\cite{morton1966computer}. PLAS is designed to preserve spatial locality by solving an assignment problem between Gaussian splats and grid positions. Although it can generate relatively coherent layouts, it relies on iterative refinement and block-wise updates, resulting in high computational cost and long sorting time. This makes PLAS less suitable for real-time or large-scale applications. In contrast, Morton ordering, also known as Z-order sorting, is computationally efficient and easy to implement. By interleaving coordinate bits, Morton curves preserve spatial proximity in a fast and deterministic manner. However, when the 1D sorted sequence is directly reshaped into 2D video frames using conventional row-major order, the resulting layout may still suffer from weak 2D neighborhood consistency and insufficient block regularity. In other words, although Morton sorting can improve 1D locality, it does not fully address the spatial continuity requirements of 2D video coding. These observations reveal a fundamental trade-off: existing methods either achieve stronger layout coherence at the cost of heavy computation, or provide efficient sorting but limited 2D spatial structure. A practical solution should therefore simultaneously maintain sorting efficiency and enhance the spatial regularity of the generated 2D maps.

To address this issue, we propose a dual-phase Morton spatial sorting framework that reorganizes Gaussian splat attributes into a coding-friendly 2D layout. The main motivation is to bridge the mismatch between the irregular structure of 3D Gaussian data and the block-based processing assumptions of conventional video codecs. Instead of directly reshaping the Gaussian sequence into the raster scan order, the proposed framework first imposes a spatially coherent ordering on the splats and then arranges the ordered samples into a 2D map with stronger neighborhood continuity. This two-step design is intended to preserve locality at both the 3D attribute level and the 2D image level, thereby improving spatial regularity, local correlation, and block consistency in the generated feature maps. As a result, the reordered attributes become much more amenable to compression by standardized video codecs, which rely heavily on local redundancy. Moreover, since the same ordering can be consistently applied to all Gaussian attributes, the proposed framework also ensures cross-channel alignment and facilitates unified coding of the entire 3DGS representation. The main contributions of this paper are summarized as follows:
\begin{itemize}
\vspace{-0.2cm}
\item We propose a dual-phase Morton sorting algorithm that transforms unstructured 3D Gaussians into structured block-wise feature maps through 1D indexing and 2D mapping, improving spatial coherence while maintaining low computational complexity.
\item We introduce a structured 2D Morton mapping table that preserves adjacency in the pixel domain, enabling seamless integration with block-based video codecs and supporting MPEG-oriented 3DGS compression workflows.
\item Experimental results demonstrate that the proposed method achieves better rate-distortion performance and lower runtime than PLAS and conventional Morton-based sorting, making it suitable for practical 3DGS applications in immersive streaming and interactive content.
\end{itemize}

\section{Method}
\vspace{-0.6em}

\subsection{Problem Formulation}
\paragraph{3DGS representation.}
A 3DGS scene is represented as a set of anisotropic Gaussian primitives, where each primitive is parameterized by both geometric and appearance-related attributes.
Specifically, for a Gaussian splat $g_i$, we denote its parameters as
\begin{equation}
g_i = \{ \mathbf{\mu}_i, \mathbf{s}_i, \mathbf{q}_i, \alpha_i, \mathbf{c}_i \},
\end{equation}
where $\mathbf{\mu}_i$ denotes the 3D mean position, $\mathbf{s}_i$ the scale, $\mathbf{q}_i$ the rotation quaternion, $\alpha_i$ the opacity, and $\mathbf{c}_i$ the spherical harmonic coefficients.

\paragraph{Objective.}
Although 3DGS provides a compact and expressive scene representation, its primitives are stored in an unstructured format that is not directly compatible with conventional video codecs.
Our goal is therefore to reorganize Gaussian attributes into structured 2D feature maps so that spatially related primitives are placed near each other in the final layout.
Formally, given a set of Gaussian splats $\mathcal{G}=\{g_i\}_{i=1}^{N}$, we seek a permutation $\pi$ and a 2D mapping $\Phi$ that minimize an adjacency-preserving loss:
\begin{equation}
\min_{\pi,\Phi} \; \mathcal{L}_{\text{adj}}(\pi,\Phi),
\end{equation}
where a smaller value indicates better preservation of spatial adjacency.
The same permutation is consistently applied to all Gaussian attributes, ensuring that the resulting layout remains aligned across attributes and amenable to video coding.

\paragraph{Design requirements.}
To achieve this goal, the proposed method is designed under three practical requirements:
\vspace{-0.5em}
\begin{itemize}\setlength{\itemsep}{0pt}\setlength{\parskip}{0pt}
    \item a unified permutation should be generated from a spatially meaningful attribute;
    \item spatial locality should be preserved both in the 1D ordering and in the 2D layout;
    \item the entire reorganization process should remain lightweight enough for large-scale GS primitives.
\end{itemize}
\vspace{-0.8em}

\subsection{Overview of the Proposed Dual-Phase Morton Sorting}
\paragraph{Pipeline description.}
As illustrated in Figure~\ref{framework_detail}, the proposed method consists of two consecutive stages.
The first stage performs Morton-based sorting on the mean positions of Gaussian splats, producing a 1D ordering that is locally coherent in 3D space.
The second stage maps this 1D sequence onto a 2D grid using a structured Morton-based layout, so that the reordered sequence also exhibits locality in the rasterized domain.
The same permutation is then applied to all remaining Gaussian attributes, and the resulting 2D feature maps are used for subsequent video coding.

\paragraph{Core idea.}
The central idea is to convert the irregular 3D arrangement of Gaussian splats into a coding-friendly 2D representation while preserving as much geometric locality as possible.
The first stage establishes a geometry-aware index order, whereas the second stage ensures that this order is reflected not only along the sequence dimension but also in the 2D spatial arrangement.
Together, these two stages reduce the mismatch between the 3D structure of the scene and the 2D block structure assumed by conventional codecs.

\subsection{Phase I: Morton-Based 1D Spatial Sorting}
\paragraph{Morton code construction.}
Morton ordering, also known as Z-ordering, is a classical space-filling curve that preserves spatial locality through bit interleaving.
Given the mean position of the $i$-th Gaussian splat, denoted as $\mathbf{\mu}_i=(x_i,y_i,z_i)$, we first quantize it to integer coordinates and then compute its Morton code by interleaving the binary bits of the three axes:
\begin{equation}
M_i = \mathcal{I}_{3D}(x_i,y_i,z_i),
\end{equation}
where $\mathcal{I}_{3D}(\cdot)$ denotes the 3D bit-interleaving operator.

In expanded form, the Morton code is obtained by concatenating the interleaved bits of the three quantized coordinates from the least significant bit (LSB) to the most significant bit (MSB):
\begin{equation}
M_i = x_{i,0} y_{i,0} z_{i,0}\, x_{i,1} y_{i,1} z_{i,1}\, \cdots\, x_{i,r} y_{i,r} z_{i,r},
\end{equation}
where $x_{i,r}$, $y_{i,r}$, and $z_{i,r}$ denote the $r$-th bits of the quantized coordinates, and $r$ indexes the bit position from the LSB to the MSB.

\paragraph{Sorting and permutation generation.}
After computing Morton codes from the mean positions of all Gaussian splats, we sort them in ascending order:
\begin{equation}
M_{\pi(1)} \leq M_{\pi(2)} \leq \cdots \leq M_{\pi(N)},
\end{equation}
where $\pi$ denotes the resulting permutation and $N$ is the number of Gaussian primitives in the current frame.
Since the sorting key is the mean position, the resulting permutation is directly driven by the spatial center of each Gaussian and therefore naturally reflects geometric proximity.
This yields a 1D ordering in which neighboring indices tend to correspond to nearby 3D locations.

\paragraph{Why Morton sorting helps.}
Compared with arbitrary indexing, Morton sorting introduces a deterministic spatial structure into the sequence without requiring iterative refinement.
This makes it particularly suitable for large GS models, where efficiency is essential.
More importantly, the resulting permutation is not used only for the mean positions: it is consistently applied to all remaining Gaussian attributes, so their values remain aligned under the same geometry-aware order.
Although these attributes are not reordered by their own semantics, the shared permutation preserves cross-attribute correspondence and provides a locality-aware basis for the subsequent 2D packing stage.

\subsection{Phase II: Structured 2D Morton Mapping}
\paragraph{Limitations of row-major reshaping.}
A straightforward way to convert the 1D sorted sequence into a 2D map is row-major reshaping.
However, this strategy preserves continuity only within each row and ignores neighborhood relations across row boundaries.
As a result, adjacent 1D indices may be placed far apart in the 2D grid, weakening local aggregation and limiting the potential gain from block-based coding.

\paragraph{2D Morton mapping principle.}
To better align the 2D layout with the spatial coherence established in Phase I, we construct a Morton-based mapping table on the 2D grid.
Let the 2D image size be $K \times K$, where $K$ is a power-of-two or a configurable grid size.
For each pixel position $(u,v)$, we compute a 2D Morton index by interleaving the binary representations of $u$ and $v$:
\begin{equation}
T(u,v) = \mathcal{I}_{2D}(u,v),
\end{equation}
where $\mathcal{I}_{2D}(\cdot)$ denotes the 2D bit-interleaving operator.
This mapping assigns consecutive 1D indices to spatially neighboring pixels, thereby producing a 2D arrangement that is more consistent with the locality-preserving order from Phase I.

\paragraph{Mapping table construction.}
Let $\Phi(\cdot)$ denote the final index mapping function. Then each sorted Gaussian index is mapped to a 2D coordinate as
\begin{equation}
\Phi(\pi(i)) = (u_i, v_i).
\end{equation}
The mapping table is constructed once and reused across frames to ensure consistent attribute layouts over time.
Such consistency is important for video coding, since it stabilizes the spatial statistics of the generated feature maps and makes the compression process more predictable.

\paragraph{Block structure and coding advantage.}
By combining 1D Morton sorting with 2D Morton reshaping, the proposed method yields feature maps with stronger local aggregation and more regular block structures.
This reduces abrupt discontinuities in the rasterized domain and increases the likelihood that nearby pixels correspond to spatially adjacent Gaussians.
As a result, the generated maps are better aligned with the assumptions exploited by conventional block-based codecs, which can improve prediction efficiency, transform compactness, and entropy coding effectiveness.

\begin{algorithm}[t]
\small
\caption{Dual-phase Morton Spatial Sorting for 3DGS Attribute Mapping}
\label{alg:dual_phase_morton}
\KwIn{Gaussian splats $\mathcal{G}=\{g_i\}_{i=1}^{N}$, where each $g_i=\{\mathbf{\mu}_i,\mathbf{s}_i,\mathbf{q}_i,\alpha_i,\mathbf{c}_i\}$; 2D map size $K \times K$}
\KwOut{Structured 2D feature maps for all Gaussian attributes}

\BlankLine
\textbf{Phase I: 1D Morton sorting}\;
\For{$i \leftarrow 1$ \KwTo $N$}{
    Quantize the 3D mean position $\mathbf{\mu}_i=(x_i,y_i,z_i)$\;
    Compute Morton code $M_i \leftarrow \mathcal{I}_{3D}(x_i,y_i,z_i)$\;
}
Sort all splats in ascending order of $\{M_i\}_{i=1}^{N}$\;
Obtain permutation index $\pi$ from the sorted order\;

\BlankLine
\textbf{Phase II: 2D Morton mapping}\;
\For{$j \leftarrow 1$ \KwTo $N$}{
    Compute 2D pixel coordinate $(u_j,v_j)$ using Morton-based mapping\;
    Store correspondence $\Phi(\pi(j)) \leftarrow (u_j,v_j)$\;
}
Construct 2D mapping table $\Phi$ from the correspondence above\;

\BlankLine
\textbf{Attribute reordering and packing}\;
\For{$j \leftarrow 1$ \KwTo $N$}{
    Reorder all attributes of $g_{\pi(j)}$ using the same permutation $\pi$\;
    Place reordered attributes into the 2D feature maps according to $\Phi(\pi(j))$\;
}
\Return Structured 2D feature maps\;
\end{algorithm}

\section{Experimental Results}

\subsection{Experimental Settings}
We conduct our experiments on the GSCodec Studio platform~\cite{11355359}, which supports three spatial sorting implementations, including PLAS, Morton, and the proposed dual-phase Morton sorting method. The test platform is equipped with an NVIDIA Tesla V100 SXM2 32GB GPU and an Intel(R) Xeon(R) Platinum 8163 CPU @ 2.50GHz with 12 cores. The experiments are performed on the 1F (static scene, single-frame) and NF (dynamic scene, multi-frame) sequences from the MPEG common testing condition~\cite{iso2026ctc}.

For performance evaluation, we use both objective quality metrics and coding efficiency metrics. Reconstructed quality is assessed using PSNR and SSIM in both RGB and YUV domains, while bitrate is reported in Mbps. We compare our method with the conventional codec HM 18.0 (HEVC reference software) and two sorting baselines, PLAS and Morton, to comprehensively evaluate both coding performance and sorting efficiency.

\subsection{Experimental Analysis}

\subsubsection{Rate-Distortion Performance}
\begin{figure}[t]
    \centering
    \vspace{-1.5cm}
    \subfloat[Rate-PSNR (RGB)]{
        \includegraphics[width=0.33\textwidth]{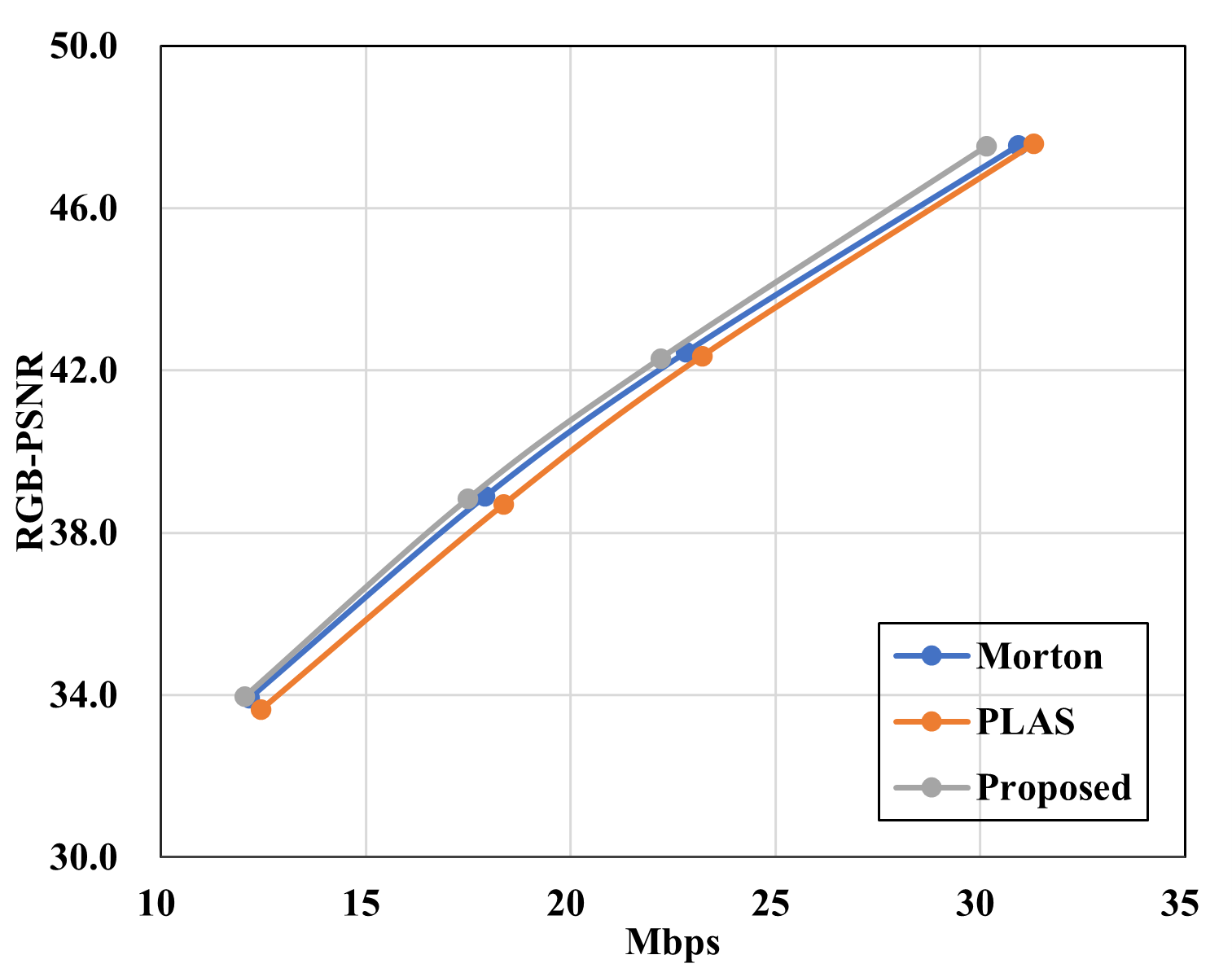}
        \label{fig:rd_rgb}
    }
    \subfloat[Rate-PSNR (YUV)]{
        \includegraphics[width=0.33\textwidth]{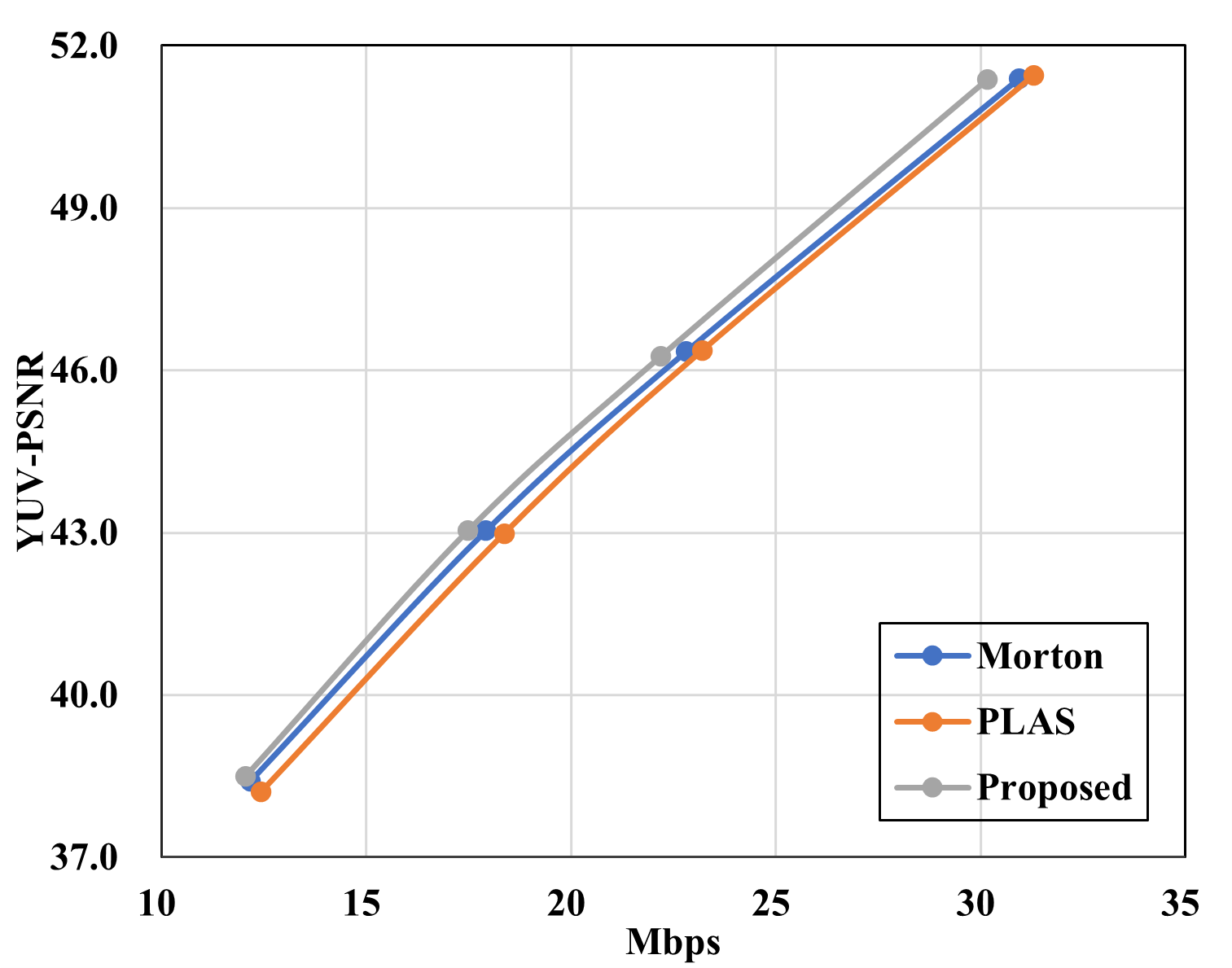}
        \label{fig:rd_yuv}
    }
    \subfloat[Rate-SSIM (YUV)]{
        \includegraphics[width=0.33\textwidth]{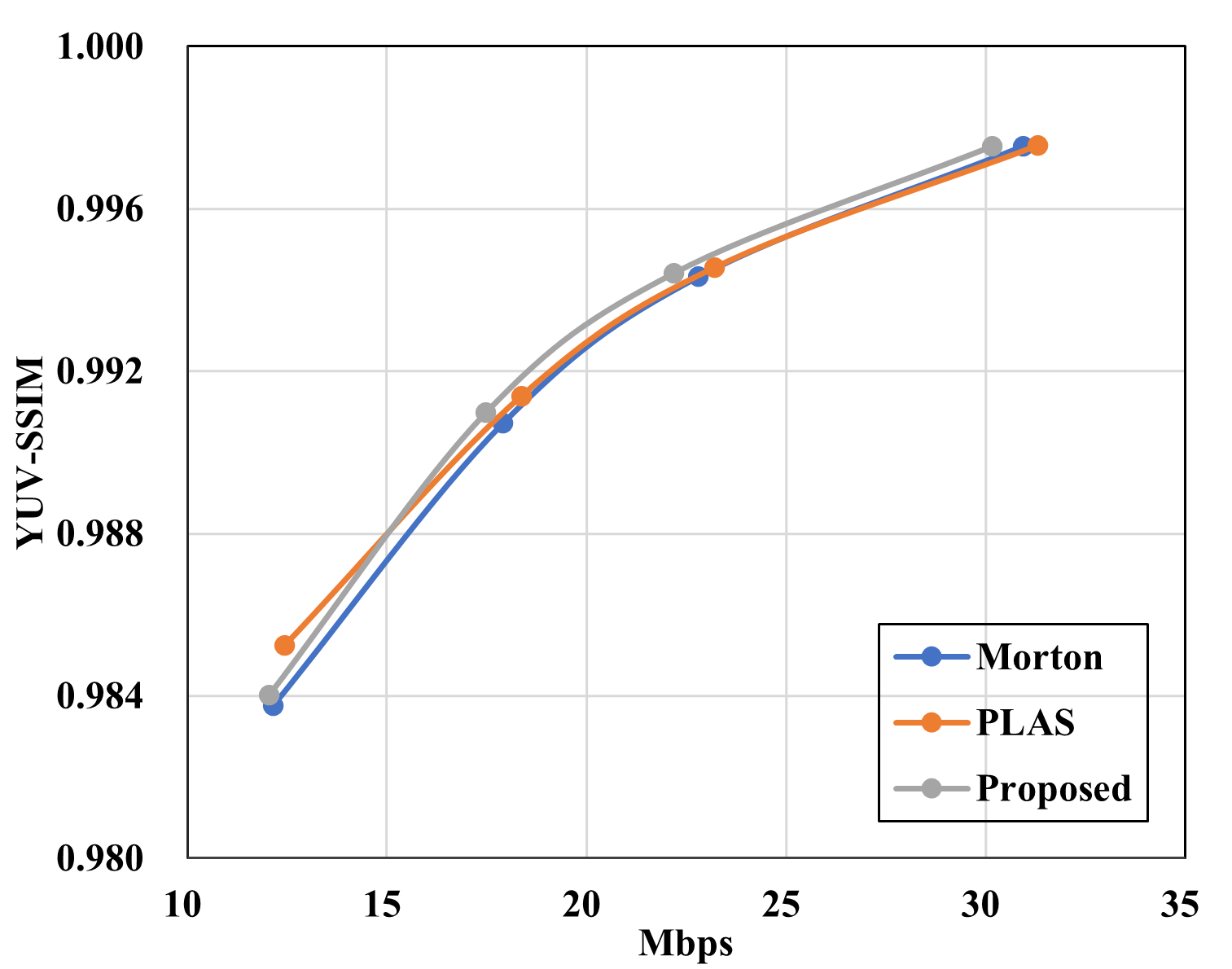}
        \label{fig:rd_ssim}
    }
    \caption{Average RD performance of different spatial sorting methods in GSCodec Studio for various 1F and NF test sequences.}
    \label{RDresult}
    \vspace{-0.4cm}
\end{figure}

As shown in Figure~\ref{RDresult}, our proposed spatial sorting method generally achieves better rate-distortion performance than PLAS and Morton across different test sequences in GSCodec Studio. 
These results suggest that the proposed sorting strategy can help preserve spatial correlation among Gaussian primitives during compression, thereby improving bitstream compactness and coding efficiency in many cases.

\begin{table*}[t]
\centering
\vspace{-0.7cm}
\caption{BD-rate comparison between Morton/PLAS and the proposed method on forward-facing and object-centric sequences.}
\label{tab:bdrate_compare}
\scriptsize
\setlength{\tabcolsep}{3.5pt}
\renewcommand{\arraystretch}{1.08}
\resizebox{\textwidth}{!}{%
\begin{tabular}{lcccccc}
\toprule
\textbf{Sequence} & \multicolumn{3}{c}{\textbf{Morton v.s. Proposed}} & \multicolumn{3}{c}{\textbf{PLAS v.s. Proposed}} \\
\cmidrule(lr){2-4} \cmidrule(lr){5-7}
 & \textbf{RGB-PSNR} & \textbf{YUV-PSNR} & \textbf{YUV-SSIM} & \textbf{RGB-PSNR} & \textbf{YUV-PSNR} & \textbf{YUV-SSIM} \\
\midrule
\multicolumn{7}{l}{\textbf{NF: Forward facing sequences}} \\
bartender\_tracked & -2.77\% & -3.31\% & -4.21\% & -0.25\% & -0.40\% & -0.46\% \\
cinema\_tracked    & -5.84\% & -6.30\% & -6.76\% & -3.33\% & -3.30\% & -3.63\% \\
breakfast\_tracked & -1.33\% & -1.51\% & -5.18\% & 2.00\% & 2.72\% & -2.60\% \\
\textbf{Average}   & \textbf{-3.31\%} & \textbf{-3.71\%} & \textbf{-5.38\%} & \textbf{-0.53\%} & \textbf{-0.33\%} & \textbf{-2.23\%} \\
\midrule
\multicolumn{7}{l}{\textbf{1F: Forward facing sequences}} \\
bartender\_semitracked & 1.86\%  & 1.02\%  & -1.21\% & 0.15\%  & -1.14\% & 1.06\% \\
cinema\_semitracked    & -3.51\% & -4.04\% & -4.87\% & -6.52\% & -6.91\% & -5.01\% \\
breakfast\_semitracked & 3.77\%  & 3.19\%  & -2.40\% & 6.06\%  & 7.87\%  & 0.44\% \\
breakfast\_untracked   & -0.31\% & -0.80\% & -2.88\% & -7.00\% & -6.86\% & -0.82\% \\
breakdance\_untracked  & -1.45\% & -1.71\% & -2.81\% & 1.54\%  & 2.09\%  & 4.47\% \\
bartender\_tracked     & -1.36\% & -1.91\% & -3.74\% & -4.44\% & -3.51\% & -4.15\% \\
cinema\_tracked        & -2.39\% & -3.19\% & -3.97\% & -8.53\% & -8.25\% & -6.24\% \\
breakfast\_tracked     & 4.43\%  & 4.33\%  & -1.52\% & 5.34\%  & 6.69\%  & 1.08\% \\
\textbf{Average}       & \textbf{0.13\%}  & \textbf{-0.39\%} & \textbf{-2.93\%} & \textbf{-1.68\%} & \textbf{-1.25\%} & \textbf{-1.15\%} \\
\midrule
\multicolumn{7}{l}{\textbf{1F: Object-centric sequences}} \\
manwithfruit\_tracked & -2.76\% & -2.74\% & -3.12\% & -4.32\% & -4.26\% & -8.24\% \\
lego\_ferrari         & -0.79\% & -0.93\% & -1.08\% & 5.87\%  & 5.81\%  & 7.47\% \\
lego\_bugatti         & -0.23\% & -0.40\% & -0.78\% & 5.22\%  & 5.20\%  & 6.46\% \\
cricket\_player       & -1.67\% & -1.79\% & -1.50\% & -1.44\% & -0.73\% & 6.74\% \\
plant                 & -0.21\% & -0.39\% & -1.37\% & 3.63\%  & 3.93\%  & 7.51\% \\
solo\_tango\_female   & -0.93\% & -1.42\% & -1.87\% & 6.76\%  & 6.58\%  & 11.45\% \\
solo\_tango\_male     & -2.50\% & -2.89\% & -1.77\% & 7.59\%  & 7.61\%  & 12.29\% \\
tango\_duo            & -0.76\% & -0.96\% & -0.91\% & 1.61\%  & 2.78\%  & 8.98\% \\
tennis\_player        & -2.18\% & -2.55\% & -2.41\% & -0.33\% & 0.08\%  & 2.89\% \\
library               & -1.46\% & -1.83\% & -2.87\% & -5.34\% & -5.90\% & -5.95\% \\
flowerdance           & 1.61\%  & 1.41\%  & 0.43\%  & -12.60\% & -12.26\% & -9.43\% \\
gymnast               & 3.76\%  & 2.63\%  & 2.16\%  & -21.73\% & -21.54\% & -15.75\% \\
\textbf{Average}      & \textbf{-0.68\%} & \textbf{-0.99\%} & \textbf{-1.26\%} & \textbf{-1.26\%} & \textbf{-1.06\%} & \textbf{2.04\%} \\
\midrule
\textbf{Total Average} & \textbf{-1.29\%} & \textbf{-1.69\%} & \textbf{-3.19\%} & \textbf{-1.15\%} & \textbf{-0.88\%} & \textbf{-0.45\%} \\
\bottomrule
\end{tabular}%
}
\end{table*}

\subsubsection{BD-Rate Savings}

Table~\ref{tab:bdrate_compare} summarizes the BD-rate savings of the proposed method over Morton and PLAS. Overall, the proposed method achieves average gains of \textbf{1.29\%}, \textbf{1.69\%}, and \textbf{3.19\%} over Morton, and \textbf{1.15\%}, \textbf{0.88\%}, and \textbf{0.45\%} over PLAS in terms of RGB-PSNR, YUV-PSNR, and YUV-SSIM, respectively. These results indicate improved rate-distortion efficiency on average across the evaluated datasets.

Nevertheless, the performance is not uniform across all sequences. Some challenging content, such as \textit{gymnast} and \textit{flowerdance}, exhibits noticeable BD-rate losses, likely due to complex motion and highly non-uniform appearance distributions, which reduce the alignment between the Morton-based ordering and the underlying compressibility structure. These results indicate that while the proposed method is effective on average, its benefit can vary depending on scene characteristics.

\subsubsection{Sorting Efficiency Comparisons}

Table~\ref{tab:sorting_time} reports the average sorting time of PLAS, Morton, and the proposed method on forward-facing and object-centric sequences from scattered Gaussian attributes to structured video maps. Overall, the proposed method substantially reduces sorting time compared with PLAS, decreasing the total average from 18,698.91 ms to 144.63 ms, which indicates a significant efficiency advantage. Compared with Morton, the proposed method incurs a moderate increase in sorting time, rising from 24.51 ms to 144.63 ms, because it performs Morton sorting twice to obtain a better trade-off between efficiency and sorting quality.

\begin{table*}[t]
\centering
\caption{Sorting efficiency comparisons (millisecond) of Morton, PLAS, and the proposed method.}
\label{tab:sorting_time}
\scriptsize
\setlength{\tabcolsep}{3.5pt}
\renewcommand{\arraystretch}{1.08}
\resizebox{\textwidth}{!}{%
\begin{tabular}{lccccccc}
\toprule
\textbf{Sequence} & \textbf{Frame} & \textbf{GOPs} & \textbf{Gaussian Number} & \textbf{Map Size} & \textbf{PLAS} & \textbf{Morton} & \textbf{Proposed} \\
\midrule
\multicolumn{8}{l}{\textbf{NF: Forward facing sequences}} \\
bartender\_tracked & 32 & 2 & 567724/585036 & 760$\times$760/768$\times$768 & 30284.78 & 71.32 & 237.54 \\
cinema\_tracked & 32 & 2 & 423249/421615 & 656$\times$656 & 21991.43 & 53.15 & 213.38 \\
breakfast\_tracked & 32 & 2 & 528154/523766 & 728$\times$728 & 27492.41 & 66.62 & 225.94 \\
\textbf{Average Sorting Time} &  &  &  &  & \textbf{26589.54} & \textbf{63.70} & \textbf{225.62} \\
\midrule
\multicolumn{8}{l}{\textbf{1F: Forward facing sequences}} \\
bartender\_semitracked & 1 & 1 & 784455 & 888$\times$888 & 20689.40 & 6.56 & 97.31 \\
cinema\_semitracked & 1 & 1 & 682712 & 832$\times$832 & 18083.88 & 5.74 & 89.23 \\
breakfast\_semitracked & 1 & 1 & 531263 & 736$\times$736 & 14759.12 & 4.46 & 86.86 \\
breakfast\_untracked & 1 & 1 & 500658 & 712$\times$712 & 12430.18 & 4.19 & 80.10 \\
breakdance\_untracked & 1 & 1 & 500000 & 712$\times$712 & 10733.41 & 4.21 & 83.09 \\
bartender\_tracked & 1 & 1 & 567724 & 760$\times$760 & 13292.44 & 4.76 & 84.92 \\
cinema\_tracked & 1 & 1 & 423249 & 656$\times$656 & 11709.94 & 3.55 & 78.80 \\
breakfast\_tracked & 1 & 1 & 528154 & 728$\times$728 & 13635.11 & 4.47 & 85.20 \\
\textbf{Average Sorting Time} &  &  &  &  & \textbf{14416.69} & \textbf{4.74} & \textbf{85.69} \\
\midrule
\multicolumn{8}{l}{\textbf{1F: Object-centric sequences}} \\
manwithfruit\_tracked & 1 & 1 & 46011 & 216$\times$216 & 4153.99 & 2.19 & 6.23 \\
lego\_ferrari & 1 & 1 & 198764 & 448$\times$448 & 4511.31 & 2.43 & 21.36 \\
lego\_bugatti & 1 & 1 & 199403 & 448$\times$448 & 4629.17 & 2.46 & 19.65 \\
cricket\_player & 1 & 1 & 196286 & 448$\times$448 & 4641.33 & 2.43 & 19.62 \\
plant & 1 & 1 & 189736 & 440$\times$440 & 3643.90 & 2.48 & 19.31 \\
solo\_tango\_female & 1 & 1 & 187896 & 440$\times$440 & 3761.05 & 2.42 & 20.70 \\
solo\_tango\_male & 1 & 1 & 183470 & 432$\times$432 & 3444.72 & 2.39 & 19.41 \\
tango\_duo & 1 & 1 & 199053 & 448$\times$448 & 4537.10 & 2.47 & 20.58 \\
tennis\_player & 1 & 1 & 192108 & 440$\times$440 & 4284.33 & 2.40 & 19.59 \\
library & 1 & 1 & 3053189 & 1752$\times$1752 & 95133.86 & 26.39 & 1112.15 \\
flowerdance & 1 & 1 & 1000000 & 1000$\times$1000 & 25437.38 & 6.55 & 95.61 \\
gymnast & 1 & 1 & 1000000 & 1000$\times$1000 & 22907.77 & 6.56 & 96.66 \\
\textbf{Average Sorting Time} &  &  &  &  & \textbf{15090.49} & \textbf{5.10} & \textbf{122.57} \\
\midrule
\textbf{Total Average Sorting Time} &  &  &  &  & \textbf{18698.91} & \textbf{24.51} & \textbf{144.63} \\
\bottomrule
\end{tabular}%
}
\end{table*}

\subsubsection{Subjective Comparisons for Different Reorganized Attribute Videos}
\begin{figure*}[t]
\centering
\vspace{-1cm}
\includegraphics[width=0.65\linewidth]{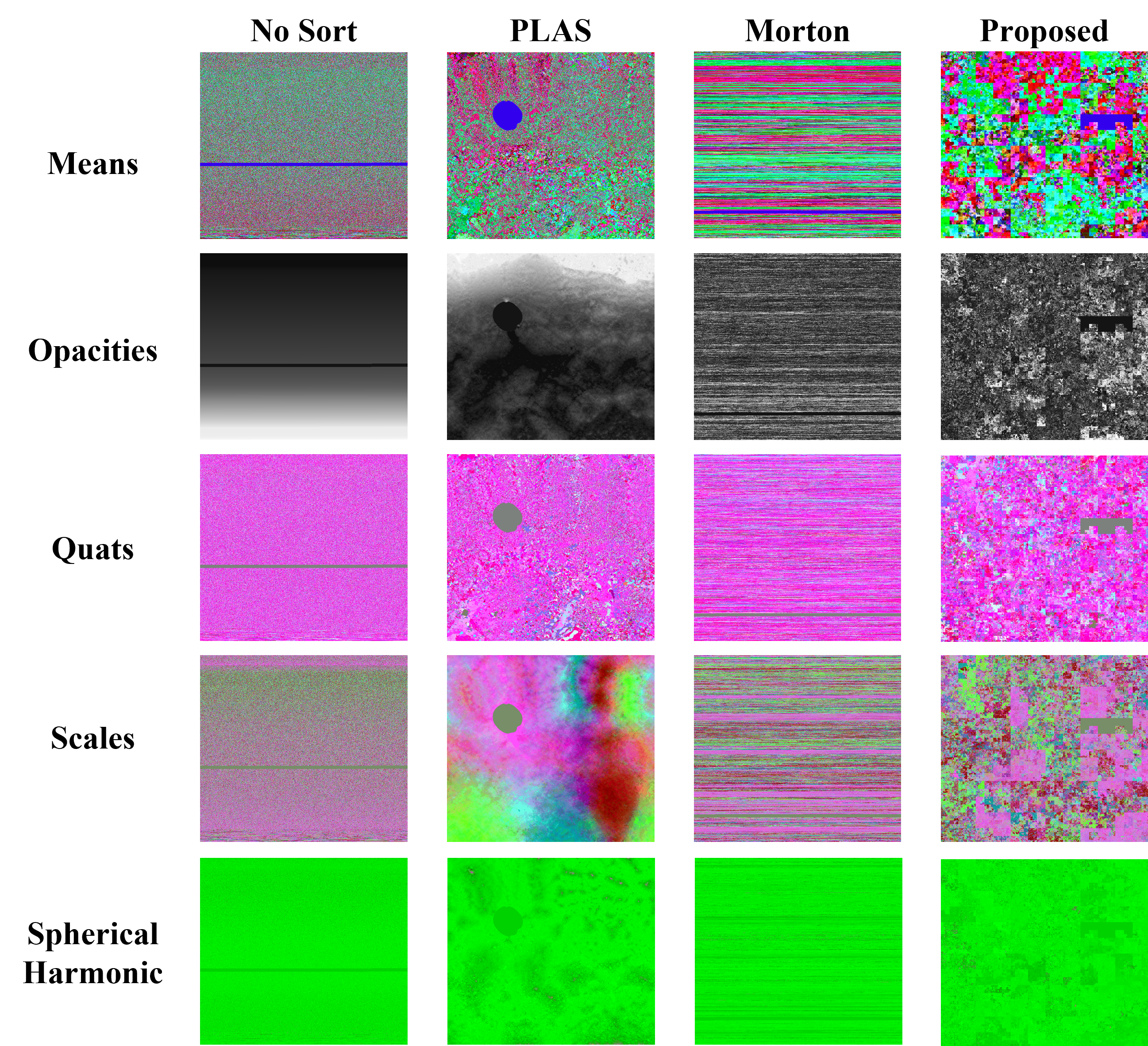}
\caption{Visualization for different attribute videos using different sorting/reorganization methods from the tracked Bartender sequence. }
\label{sorting_pattern}
\vspace{-0.4cm}
\end{figure*}

As shown in Figure~\ref{sorting_pattern}, the proposed sorting algorithm produces more compact and locally clustered GS attribute videos, which are better aligned with block-based video coding. In contrast, PLAS yields a less compact layout, and Morton sorting still leaves a more scattered arrangement. Overall, the proposed scheme generates more coding-friendly attribute videos and may improve compression performance.

\section{Conclusion}
This paper proposes a dual-phase Morton sorting algorithm for efficient 3D Gaussian Splatting coding. By combining Morton-based 1D indexing with a structured 2D mapping table, the proposed method reorganizes unstructured Gaussian attributes into block-wise feature maps with improved spatial coherence and local correlation. This structured representation is well suited for conventional block-based video codecs, enabling effective compression while maintaining compatibility with existing standards. Experimental results demonstrate that the proposed approach achieves a favorable trade-off between coding performance and computational complexity, with improved rate-distortion performance and reduced runtime compared with existing sorting strategies. These results suggest that dual-phase Morton sorting is a practical and scalable solution for 3DGS compression, and it provides a promising direction for standards-compliant representation and coding of Gaussian-based scene data.

\vspace{-0.7em}
\section*{References}
\vspace{-0.6em}
\bibliographystyle{IEEEtran}
\bibliography{main}
\vspace{-0.8em}
\end{document}